# Exploratory analysis of text duplication in peer-review reveals peer-review fraud and paper mills

Adam Day

*SAGE Publishing, 1 Oliver's Yard, London, EC1Y 1SP*

adam.day@sagepub.co.uk

## Abstract

Comments received from referees during peer-review were analysed to determine the rates of duplication and partial duplication. It is very unusual for 2 different referees to submit identical comments, so the rare cases where this happens are of interest. In some cases, it appears that paper-mills create fake referee accounts and use them to submit fake peer-review reports. These include comments that are copied and pasted across multiple reviews. Searching for duplication in referee comments is therefore an effective method to search for misconduct generally, since the forms of misconduct committed by paper-mills go beyond peer-review fraud. These search methods allow the automatic detection of misconduct candidates which may then be investigated carefully to confirm if misconduct has indeed taken place. There are innocent reasons why referees might share template reports, so these methods are not intended to automatically diagnose misconduct.

## Introduction

In recent years, there has been substantial growth in the volume of published research (STM Association, 2021). At the same time, reports of research misconduct have grown rapidly (Ivan Oransky, 2021) (Elisabeth M. Bik, 2016).

Peer-review fraud is defined here as any attempt to dishonestly influence the peer-review process. Examples of peer-review fraud include:

- An author creating a fake name and fake email address, then recommending that fake name and email address as a potential referee for their paper. If the journal invites that fake referee to review the paper, then the author can review their own paper using the fake account they created.

- A group of individuals colluding to give each-other's work lenient treatment in peer-review. This is sometimes referred to as a 'peer-review ring'.
- A 'paper mill' (an organisation which creates fake research papers and sells authorship on those fake papers) submits a paper on behalf of an author using a fake email address. If the fake author account is not identified, then the journal may invite that account to review papers.
- Or a paper mill might simply create a fake reviewer account and recommend that account as a reviewer of a paper.

Cases of peer-review fraud have been identified in the past and described in numerous papers (Misra, Ravindran, & Agarwal, 2018) (Dadkhah, Kahani, & Borchardt, 2017) (Cohen, et al., 2016) (Kamali, Abadi, & Rahimi, 2020) as well as journalistic outlets (Oransky, retractionwatch.com, 2014) (Marcus, 2013) (Oransky, retractionwatch.com, 2012) (Ferguson, 2014) (Oransky, retractionwatch.com, 2014) (Fountain, 2014) (Normile, 2017). However, there is little research on automated methods to detect or prevent this form of fraud, e.g. (Rivera & Teixeira da Silva, 2021).

In this study, we take a dataset of referees' comments submitted to SAGE Publishing during peer-review and look for duplication of text in reports submitted by 2 or more reviewers. Duplication is likely the result of multiple referee-accounts using the same template for comments. The use of templates, on its own, is not strong evidence of misconduct. However, in some cases we find several referee accounts using a number of similar templates and we also find referees using the same template as the author of the paper under review. In these cases, the evidence of misconduct is stronger.

## Data protection

Referees' comments are considered to be confidential and potentially constitute personally-identifiable data. As such, they are considered sensitive under SAGE Publishing's data protection policies (SAGE, 2021) and are handled in compliance with data protection legislation such as the European Union's General Data Protection Regulation (GDPR) (European Commission, 2021). For this reason, data and code for this study will not be made available.

## Methods

### *Data acquisition*

A dataset was retrieved from the ScholarOne peer-review management system (Clarivate, 2021). This dataset included the text of each referee's comments for each article peer-reviewed by 23 SAGE Publishing journals. Each referee-report includes 2 text fields: one field is for comments to the authors, and the other field is for comments to the journal-editors.

### *Data preparation*

The data was arranged into a pandas (pandas, 2021) dataframe using Python. Personally-identifiable data, such as email addresses, were removed or pseudonymised. Each row of the dataframe included the following fields:

- Unique article identifier
- Referee's unique identifier (pseudonymised)
- The referee's comment on the article
- An indicator showing whether the comment was to the authors or to the editors.

Several rows of data were filtered out:

- 4 journals were found to offer referees a set of template questions. All rows of data relating to these journals were filtered out because the template questions were often copied into the referees' comments.
- Very short comments below 150 characters in length were also removed.
- Comments for the first round of review for each article were retained. Comments for all other rounds of review were removed.
- Negative comments recommending rejection were removed.

Minimal data-cleaning was performed on comments.

- Extraneous whitespace was removed from the comments and accented characters were replaced with non-accented characters using the Gensim (Rehurek, 2021) module's `strip_multiple_whitespaces` and `deaccent` functions.

A second dataset was prepared based on the first. Each referee-comment was split into sentences using nltk's `sent_tokenize` function (nltk.org, 2021) and a new dataframe was created with 1 row per sentence.

| | |
|---|---|
| Number of comments | 218204 |
| Number of articles comments refer to | 66815 |
| Number of journals in dataset | 19 |
| Number of unique referee accounts | 62794 |
| Number of sentences | 5508344 |

*Table 1 Descriptive statistics of dataset used in this study*

### Similar document searches

A number of searches were carried out to find pairs of referee comments which were duplicates, or similar.

#### *- Search 1: exact duplicate comments*

A very simple check was performed to find which referee-comments were exact duplicates.

Additional pre-processing steps were applied:

- Punctuation removed
- Numeric characters removed
- Short comments below 20 words in length removed (there was some obvious convergence of short pieces of text. E.g. the following is a common comment to editors: "Thank you for the opportunity to review this paper.".)

This check can be performed on this dataset in less than 1 second using the pandas `value_counts()` function.

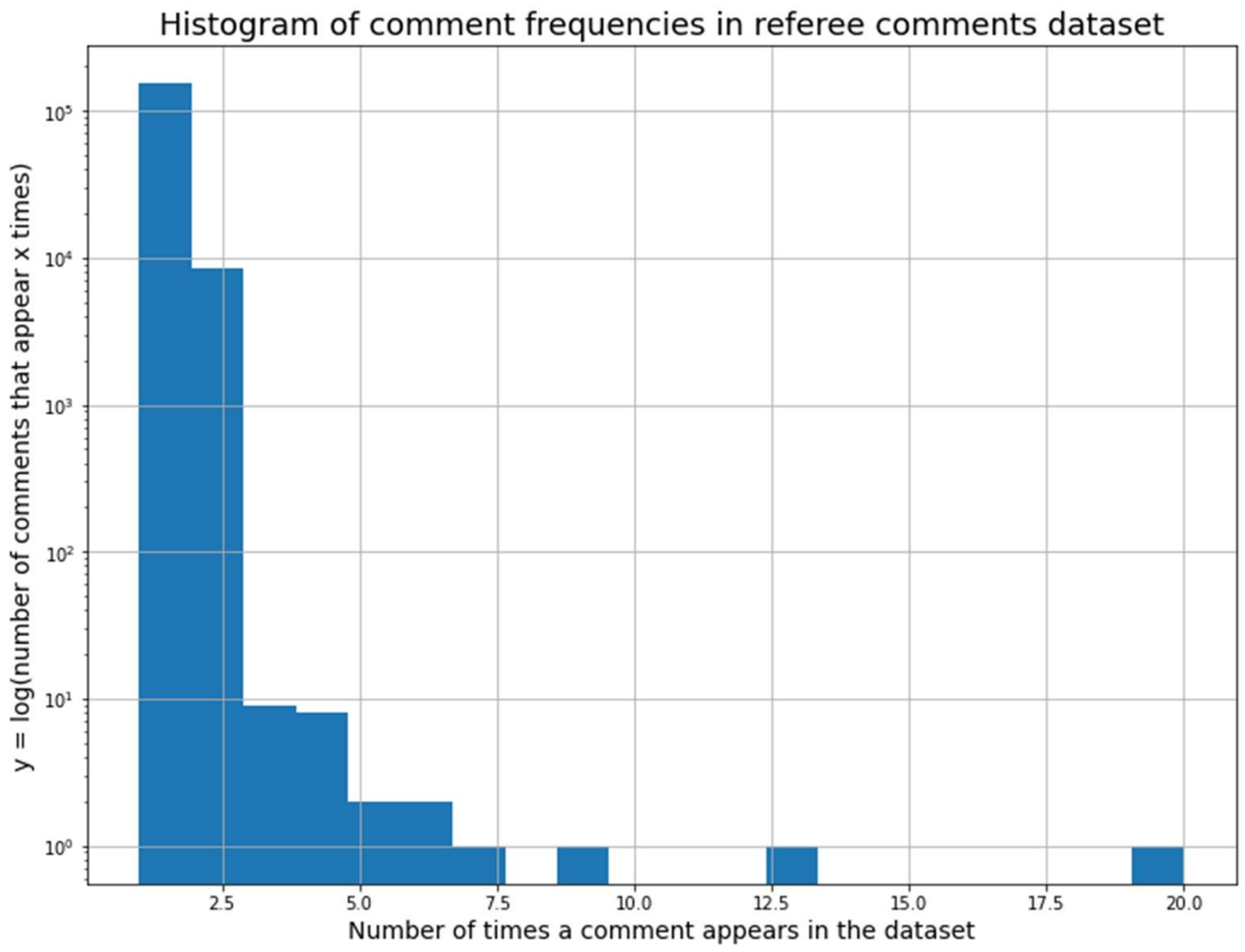

*Figure 1 Most comments appear only once or twice. Exact duplicate comments are very rare and usually the result of an individual repeatedly using a template – or pasting the same comment in to both the 'comment to authors' and the 'comment to editors' fields.*

To ensure that duplicate comments were not submitted by the same referee account, a `for` loop was created to check the referee accounts' unique ids for every duplicate pair found.

This search revealed a number of unique referee accounts which had submitted identical comments. This included:

- One large cluster of referee accounts which had submitted several duplicate comments.
- A few *pairs* of referee accounts which had submitted duplicate comments.

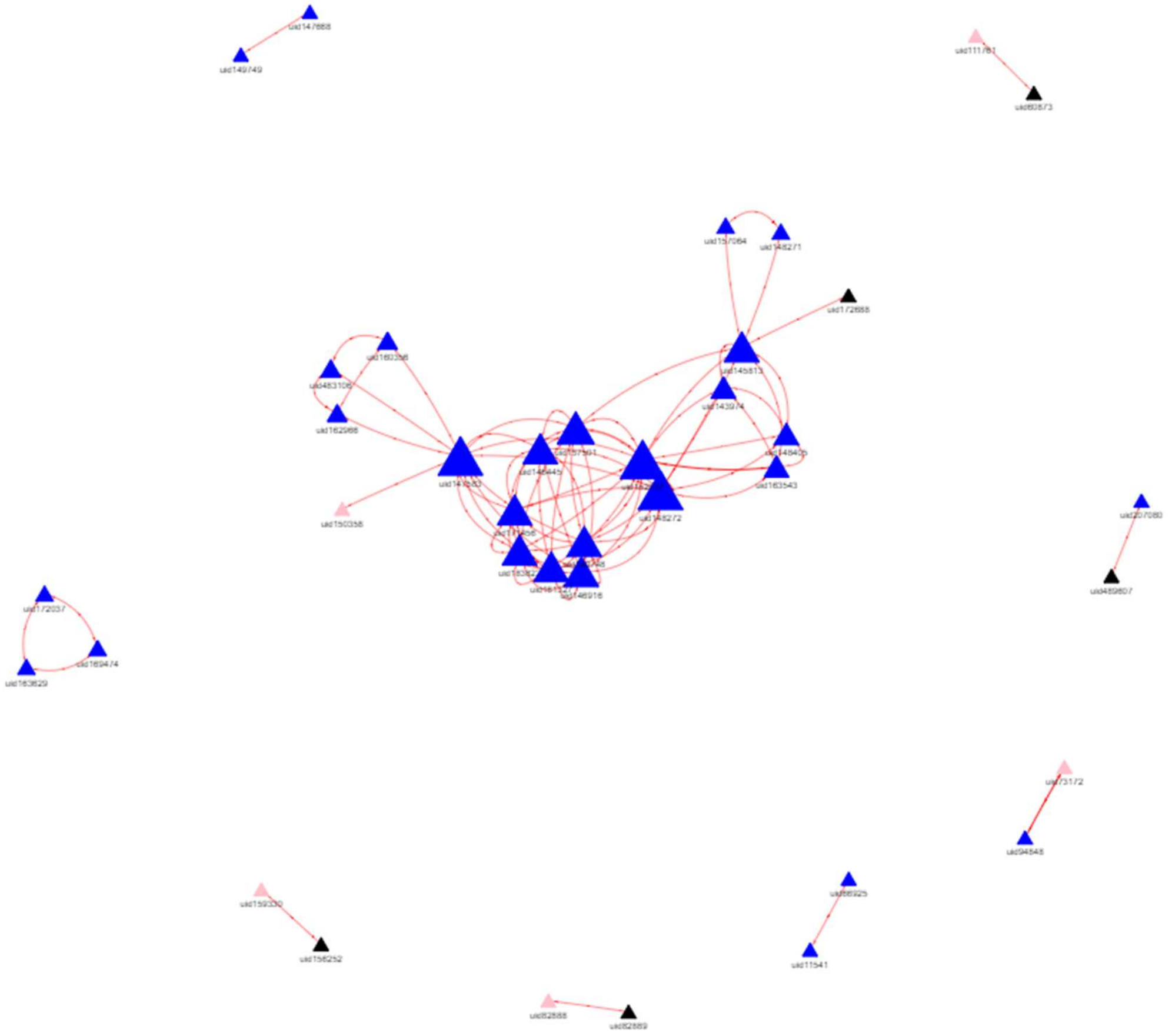

*Figure 2 Triangle-nodes represent referee-accounts blue triangles have never appeared as authors. Black triangles have been lead authors and pink triangles have been co-authors. Red lines indicate that 2 referee-accounts have submitted identical comments. The thickness of the lines shows the number of duplicated comments 2 referees have both submitted, the size of the triangle is proportional to the number of duplicated comments each referee has submitted. Each referee has a unique id of the form 'UIDxxxxx'. The UIDs serve as an ad hoc pseudonymization of the referee's identity.*

It's possible that, occasionally, referees write identical reports by chance, so the probability of these *pairs* representing deliberate duplication (and therefore peer-review fraud) is much lower than the large cluster. On the other hand, there is very little chance that we would see a large cluster of referee accounts all producing the same comment several times without organised and deliberate sharing of comments. In this case, a spot-check of the ScholarOne records for articles reviewed by these accounts showed cases where manuscripts were rejected when journal editors became suspicious that the manuscripts were submitted by a paper-mill.

We therefore make the assumption that the large cluster represents a paper-mill and that the referee accounts in this cluster are fake accounts controlled by that paper-mill.

*- Search 2: proportion of sentences overlapping between 2 comments*

Each sentence in each referee's comments was compared with each sentence in each other referee's comments. Jaccard similarity was calculated for each pair of comments, where we define Jaccard similarity as:

*Equation 1*

$$J_{ab} = \frac{\text{number of unique duplicated sentences in comment a and comment b}}{\text{total number of unique sentences in comment a and comment b}}$$

Pairs of referee comments with Jaccard similarity >0.5 are very rare.

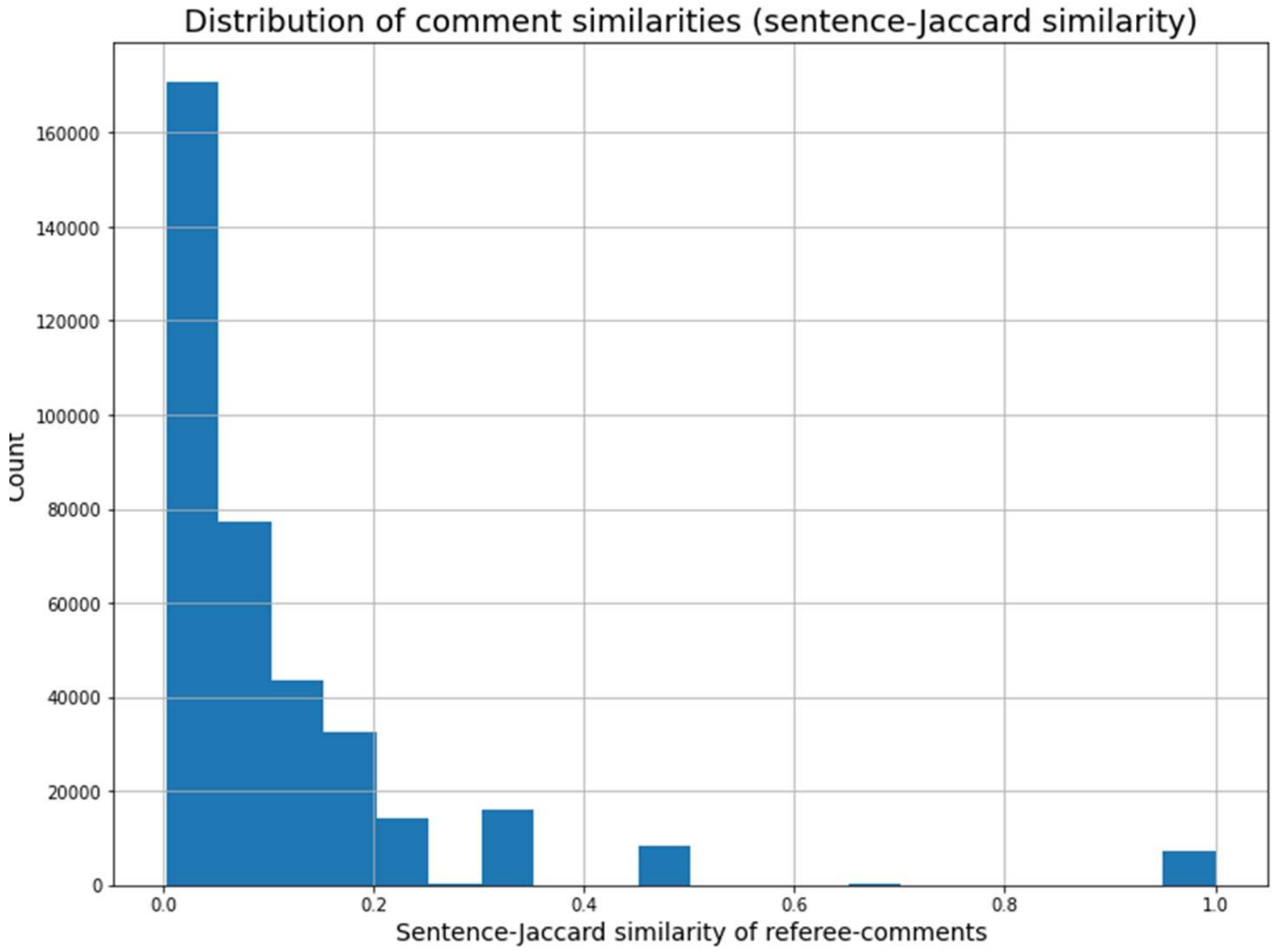

*Figure 3 Distribution of Jaccard similarities where Jaccard similarity is defined in equation 1*

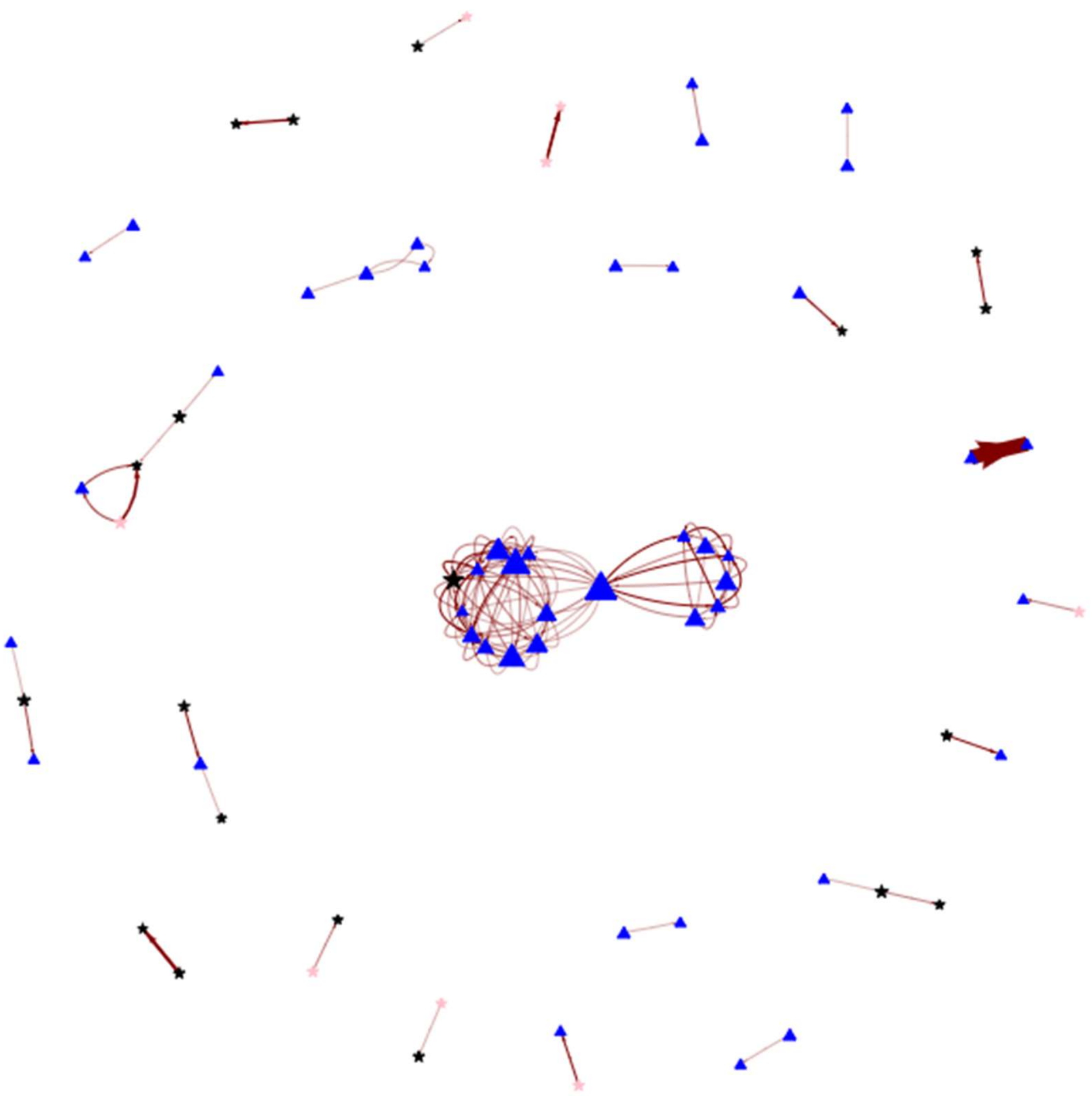

*Figure 4 Links showing where a high proportion of sentences were duplicated between comments submitted by 2 referee accounts.*

### Search 3: Locality SensitiveHashing with Datasketch

Previous searches used ‘brute force’ methods where, essentially, all referee comments are compared with each other. This is slow, so it is preferable to perform comparisons on a smaller number of pairs of documents. We achieve this by loading our referee comments into an index which allows us to rapidly retrieve small numbers of similar documents.

The MinHash Locality Sensitive Hashing (MinHash LSH) process, for detecting near-duplicate text strings, works as follows.

- Comments are broken down into ‘shingles’, which are overlapping substrings. For example: if shingle-size is set to 5, the sentence ‘the cat sat on the mat’ contains the following 5-character-length shingles: {' cat ', ' on t', ' sat ', ' the ', 'at on', 'at sa', 'cat s', 'e cat', 'e mat', 'he ca', 'he ma', 'n the', 'on th', 'sat o', 't on ', 't sat', 'the c', 'the m'}

- The shingles are indexed into clusters using the MinHash LSH process with the Python Datasketch package (Zhu E. (., 2021).
- Then each document was broken into shingles in the same way and the LSH index was used to find similar documents containing the same shingles in the LSH index.
- This method approximates the results of a brute-force search for strings with a high shingle-similarity. I.e. a high Jaccard similarity for shingles:

*Equation 2 Shingle similarity - or Jaccard similarity for strings broken into shingles.*

$$J_{ab} = \frac{\text{number of unique duplicated shingles in comment a and comment b}}{\text{total number of unique shingles in comment a and comment b}}$$

A more detailed description of the MinHash LSH algorithm is found on its author's website (Zhu E. , 2021).

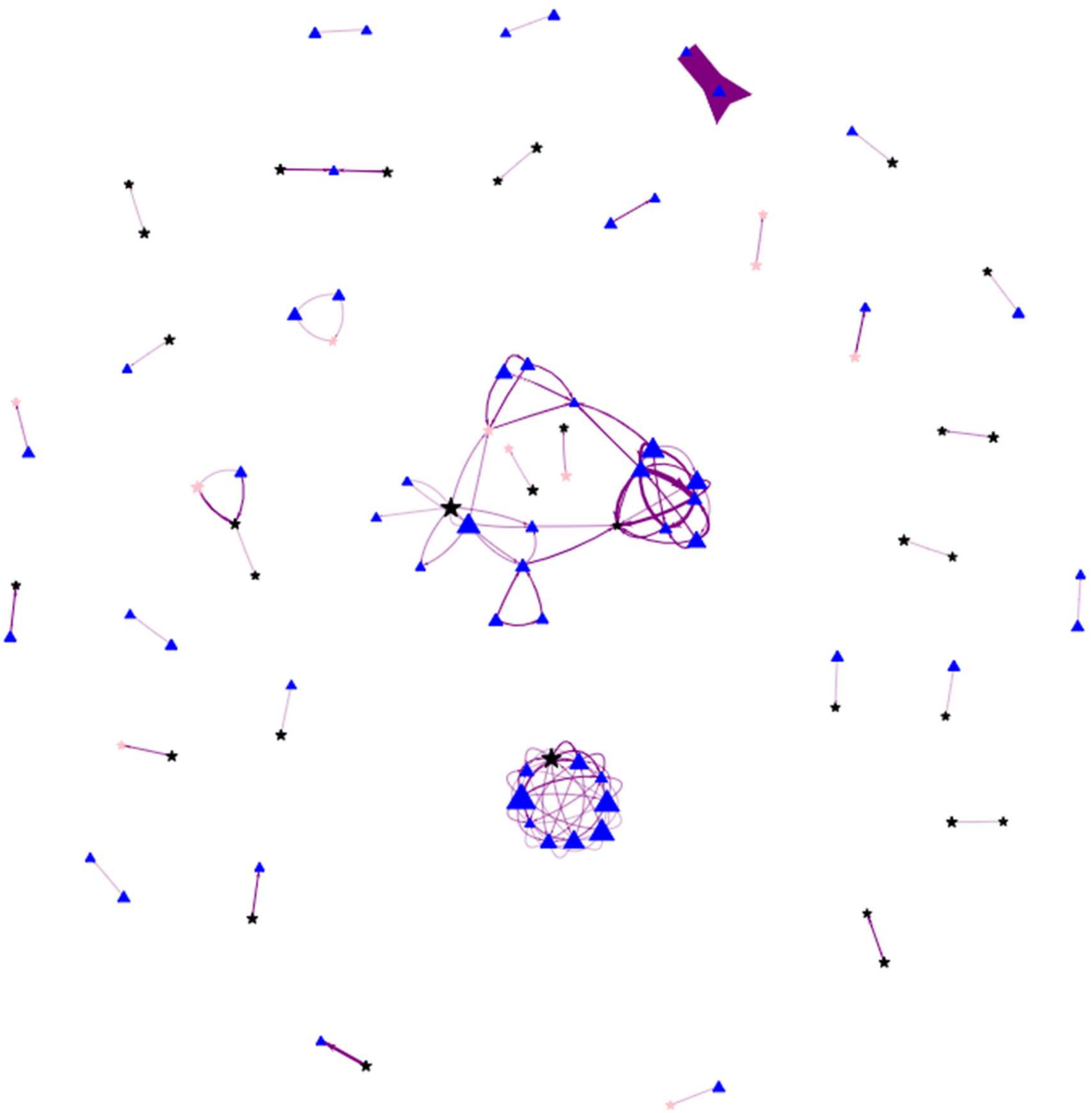

*Figure 5 Links showing where 2 referee accounts have produced similar comments measured with minHash LSH*

### *Search 4: Elasticsearch with rapidfuzz*

Elasticsearch is a highly scalable graph database with built-in text-search capabilities. Again, this allows the retrieval of similar documents to a query document and avoids the need to perform a brute-force search.

- Each comment was loaded into Elasticsearch.
- Each comment was searched-for in Elasticsearch. This yielded a list of most-similar comments for each comment.
- Comments by the same referee were removed from the results.
- Comments were compared using a number of textual similarity metrics:

- `rapidfuzz` `partial_ratio`
- `rapidfuzz` `ratio`
- `rapidfuzz` `token_sort_ratio`
- Sentence_jaccard similarity as described in Search 2
- Shingle similarity as described in Search 3

- In each case, comments were retained for the top 0.1% of results. Aside: visual inspection of the distributions below is an alternative way to choose a cutoff for each particular metric.

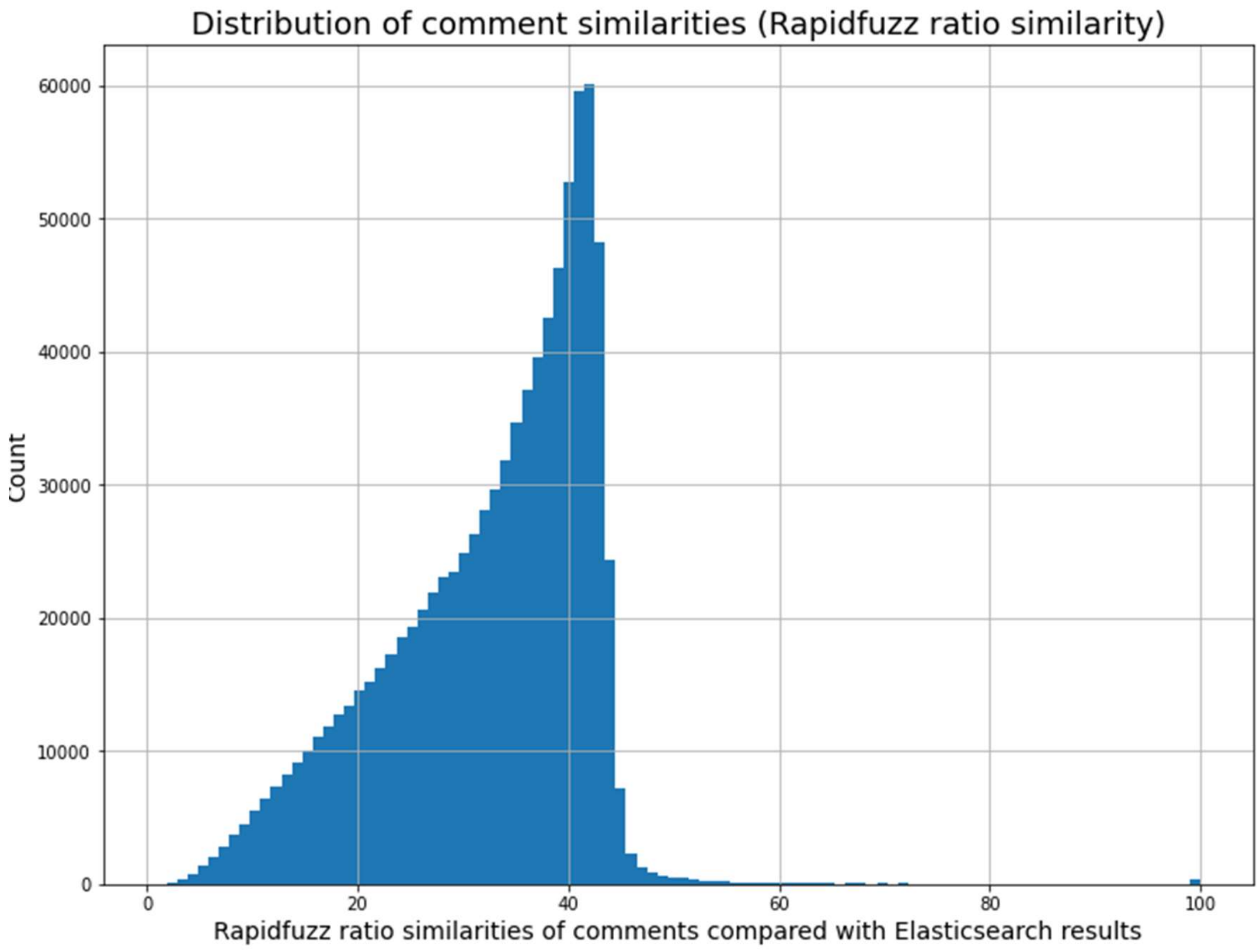

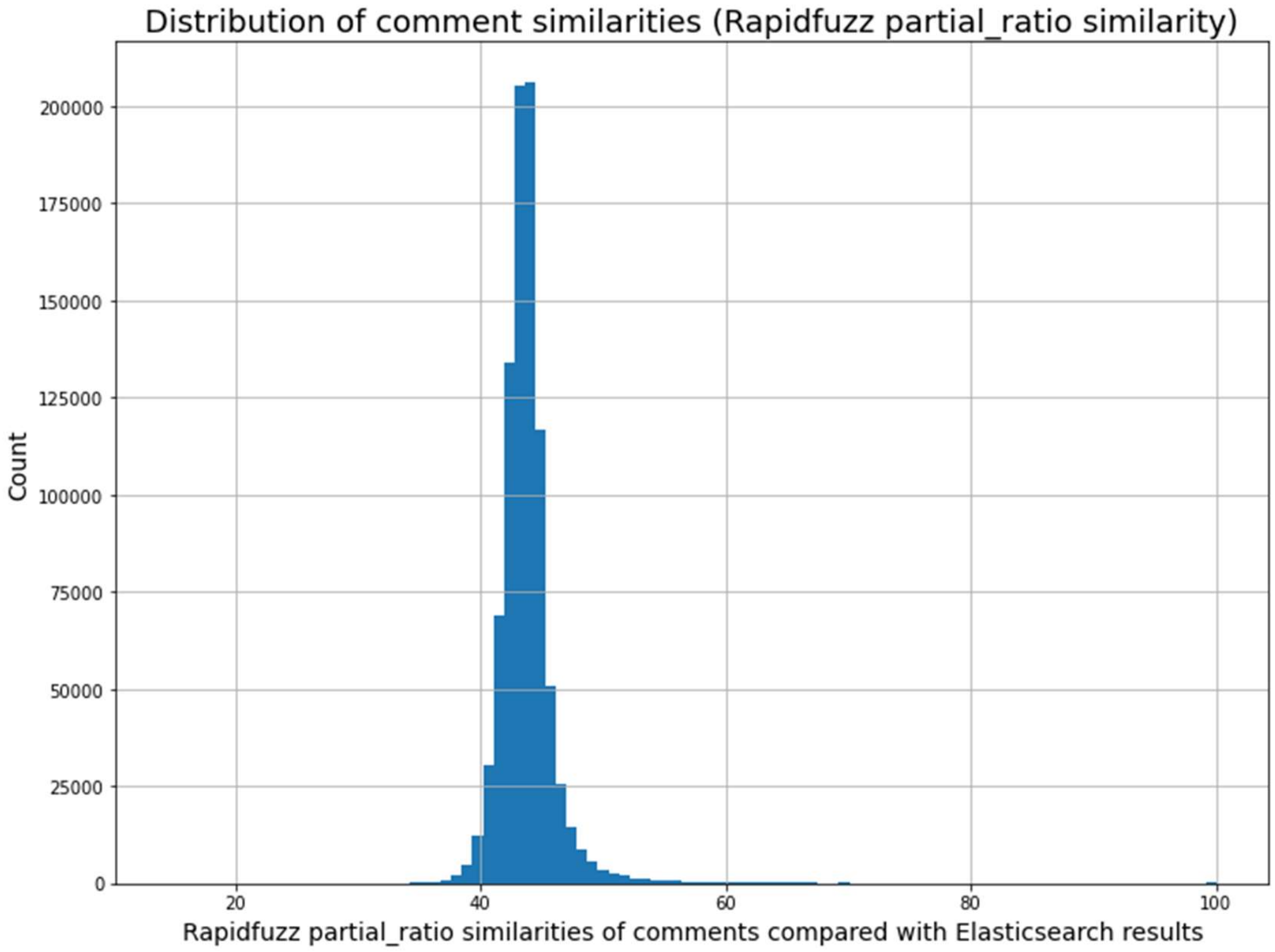
Distribution of comment similarities (Rapidfuzz partial_ratio similarity)
Count
Rapidfuzz partial_ratio similarities of comments compared with Elasticsearch results
200000
175000
150000
125000
100000
75000
50000
25000
0
20
40
60
80
100

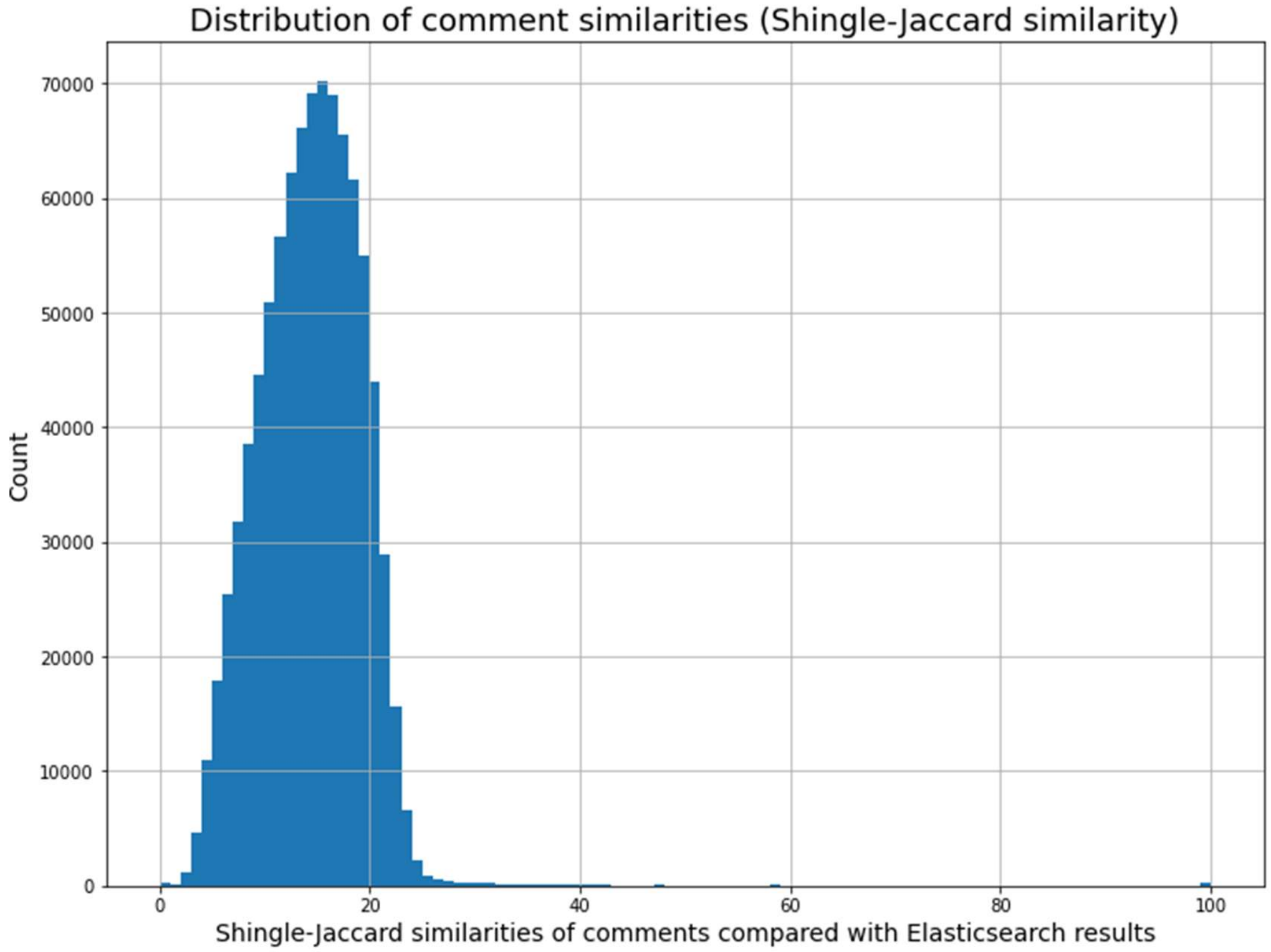
Distribution of comment similarities (Shingle-Jaccard similarity)
Count
Shingle-Jaccard similarities of comments compared with Elasticsearch results
70000
60000
50000
40000
30000
20000
10000
0
0
20
40
60
80
100

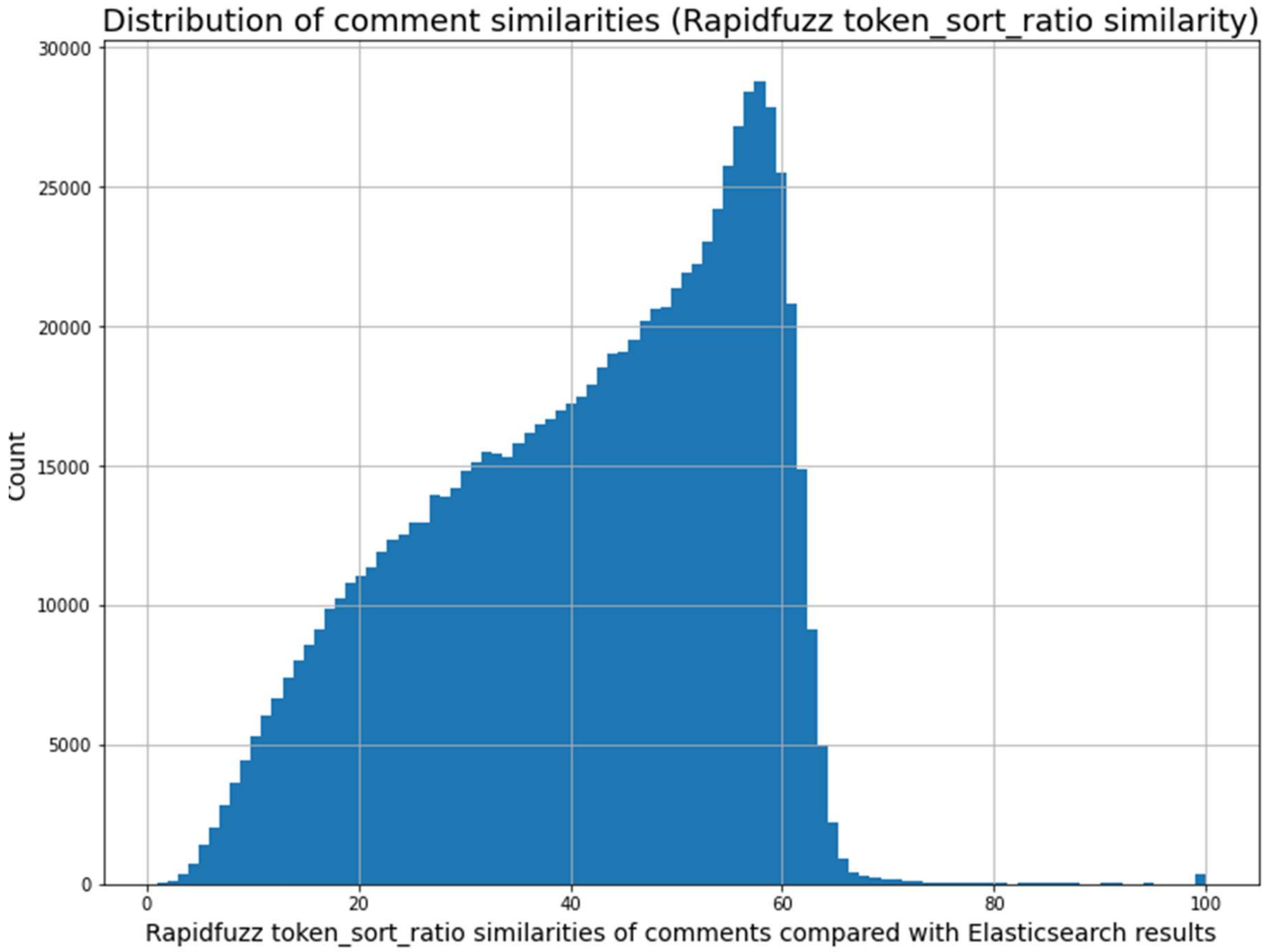
Distribution of comment similarities (Rapidfuzz token_sort_ratio similarity)
Count
Rapidfuzz token_sort_ratio similarities of comments compared with Elasticsearch results

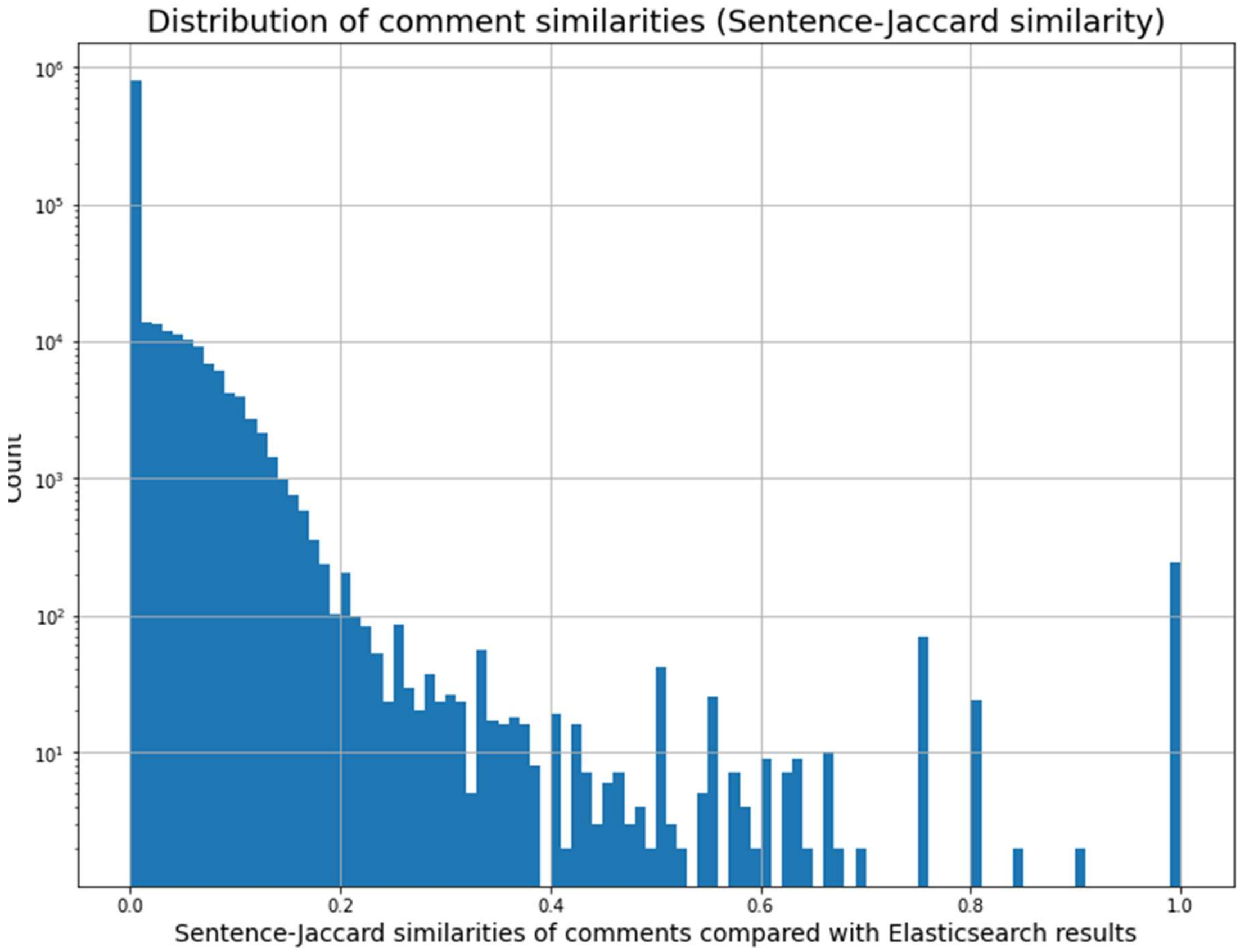
Distribution of comment similarities (Sentence-Jaccard similarity)
Count
Sentence-Jaccard similarities of comments compared with Elasticsearch results

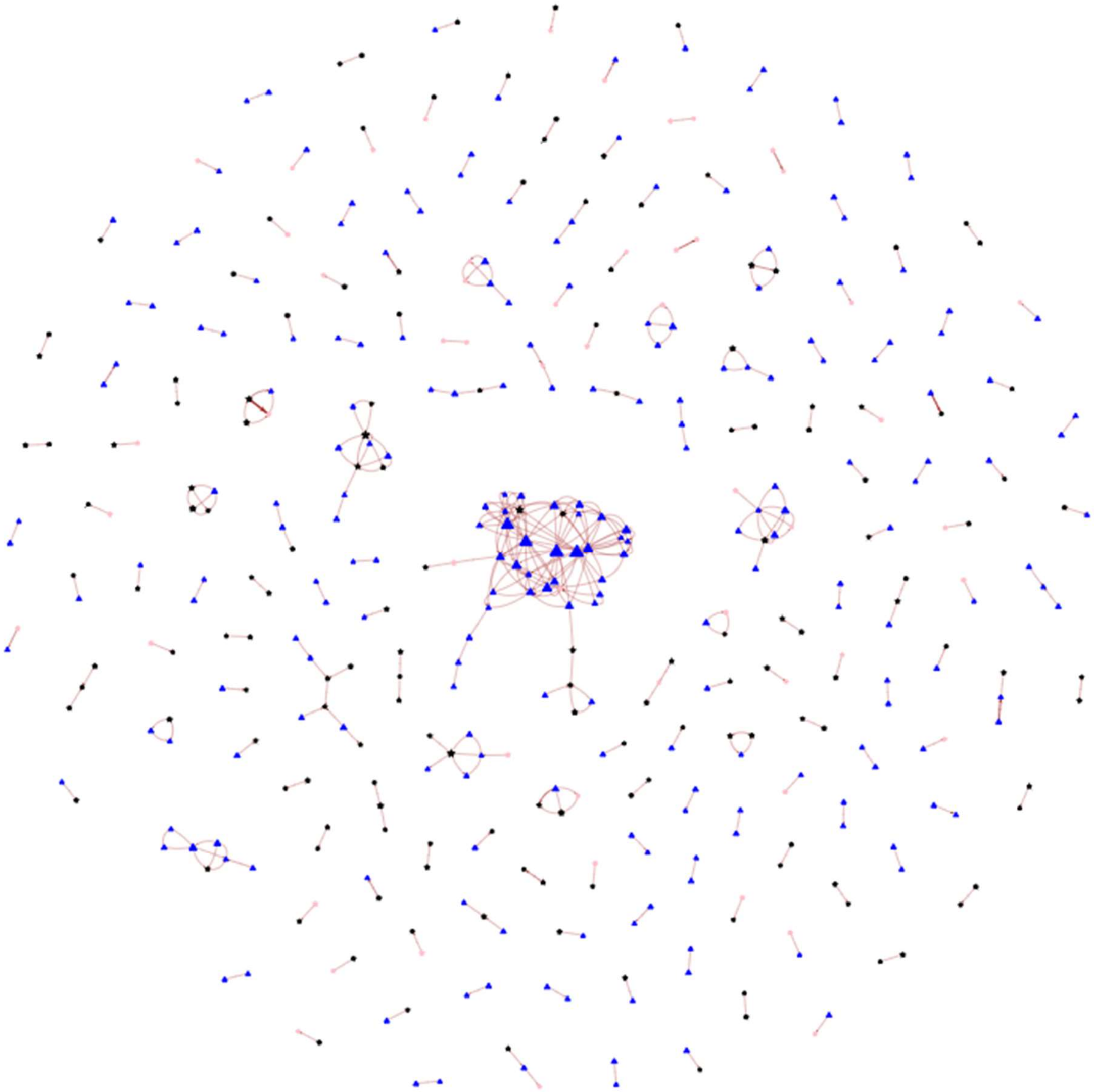

*Figure 6 Referee accounts which have produced similar reports found using elasticsearch. This method produced a lot more 'noise' i.e. pairs of accounts as opposed to clusters.*

## Template searches

We assume that, when a referee is using *template comments,* rather than copy/pasting a whole comment, it is more likely that a few sentences are copied and pasted and then the rest of the comment is adapted to the particular manuscript in-hand. This means that, rather than looking for similar comments, more duplication might be found by looking for smaller overlapping pieces of text. A number of searches were carried out to find cases where small amounts of very unique text had been used by more than one referee account.

*- Search 5: common duplicate sentences with errors*

A number of very common duplicated sentences were found in the data which contained typographical errors or linguistic mistakes. It is highly unlikely that 2 or more independent referees would coincidentally produce the same sentence *containing the same mistake*.

A table of counts was made showing which sentences have been duplicated by different referees the most.

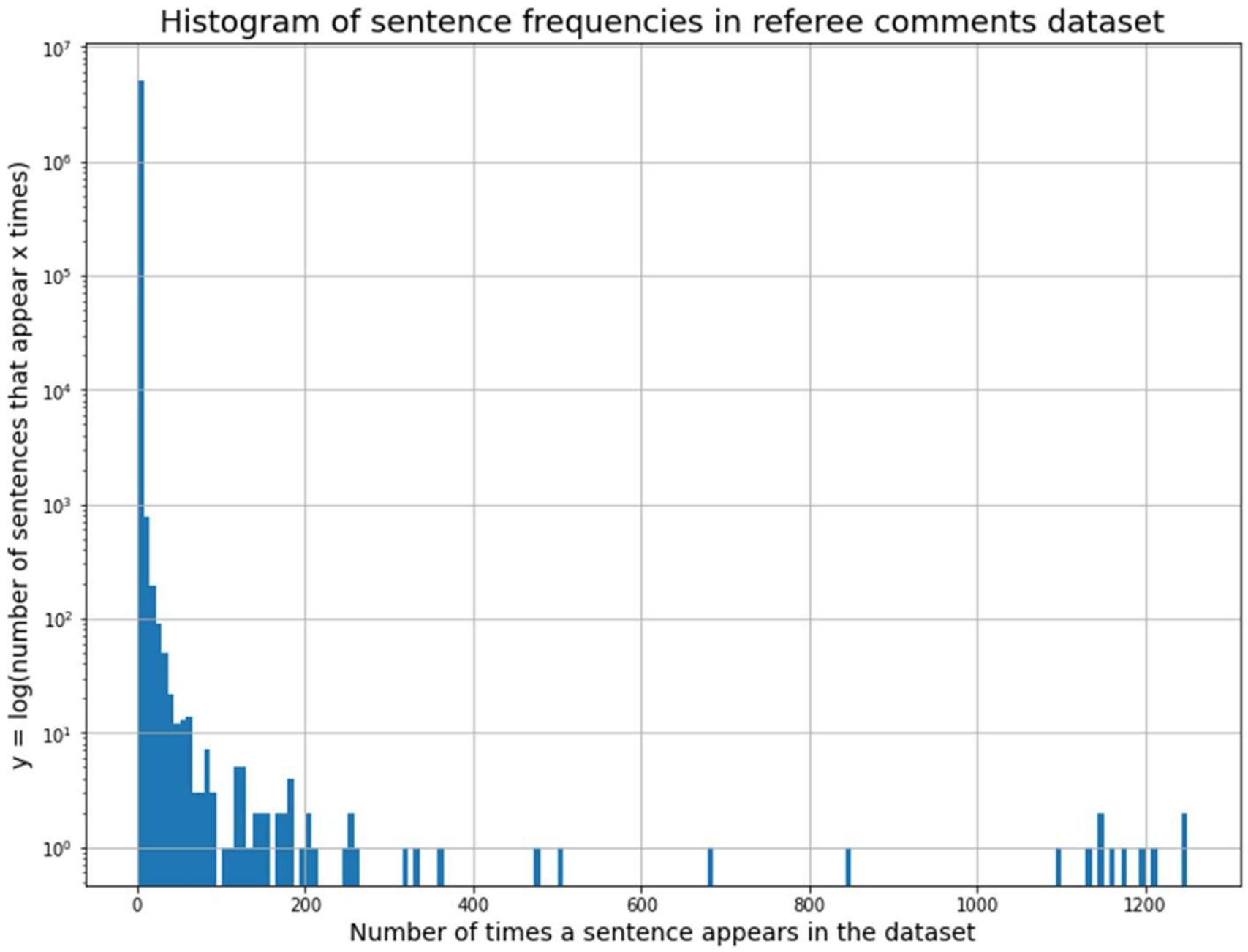

*Table 2 Note the log-transformation of the y-axis. It is clear that most sentences only appear once in the dataset. Sentences which appear a very large number of times turn out to be journal-template questions.*

There are a number of innocent scenarios where sentences are duplicated.

- Convergences: where 2 referees happen to write the same thing coincidentally. E.g.: "This is an interesting paper."
- Special collection titles: where 2 referees both wrote identical sentences like: "this paper has been submitted to [special collection title]"
- Subheadings: sometimes there are bullet-points in the instructions given to referees by a particular journal. E.g. "What is the motivation of this paper?" These subheadings are often copy/pasted into the referee-comments.
- References: referees often paste references into referee comments recommending that the authors cite a particular paper. We occasionally find different referees pasting the same reference into their referee-comments.

These scenarios are significantly more-common than the misconduct cases under study. Highly frequent sentences with obvious mistakes from the dataset were selected manually. Referee accounts which produced those errors were added to the search results.

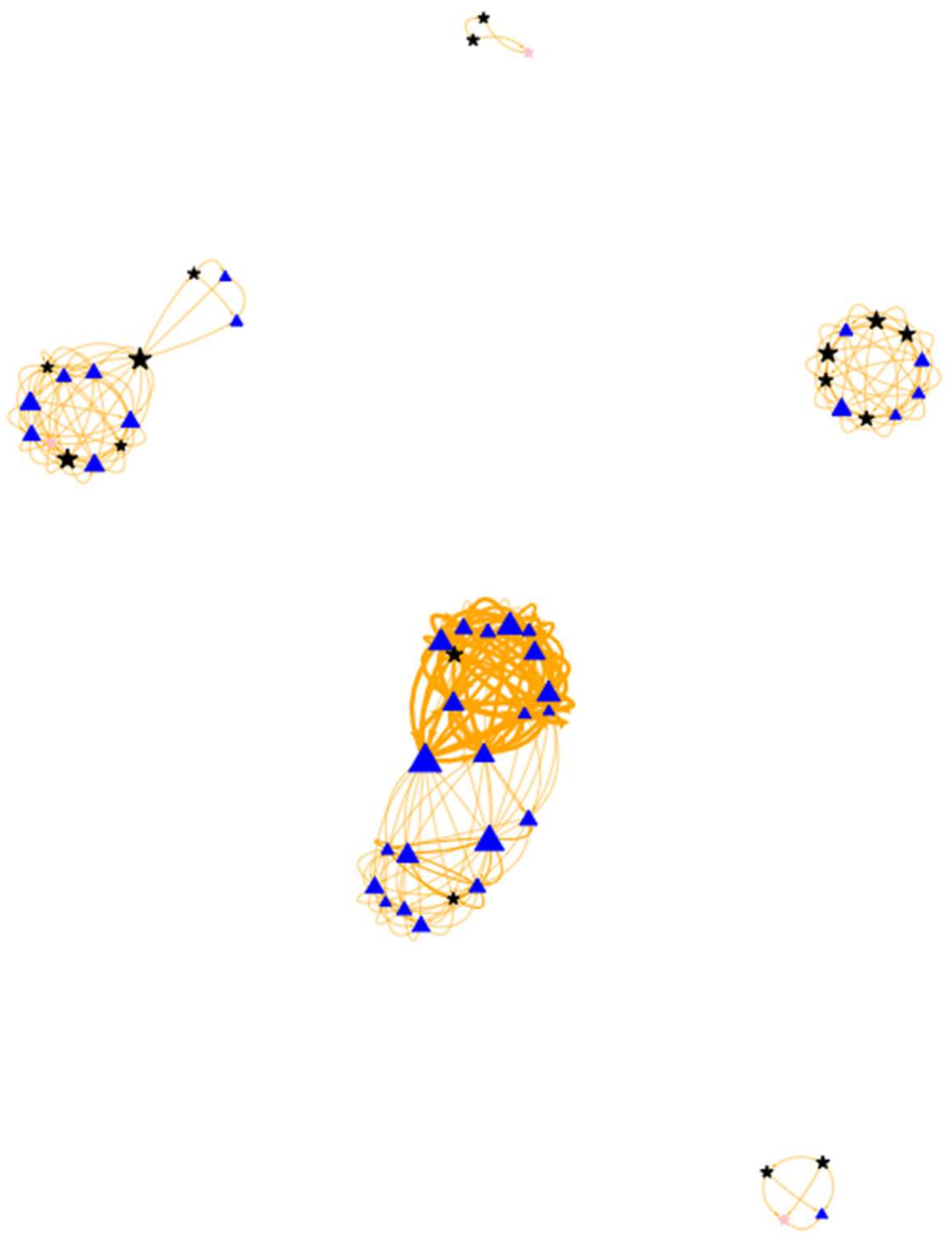

*Figure 7 Links showing where 2 referee-accounts submitted comments containing the same linguistic or typographical error.*

*Search 6: Elasticsearch sentences*

This search was aimed at finding cases where there were weak links between the referee accounts identified in previous searches and other accounts in the dataset.

In this search,

- All sentences from all referee-comments were loaded into an Elasticsearch index.
- A set of accounts (specifically, the accounts identified in previous searches) was created.
- Each sentence written by those accounts was searched-for in the Elasticsearch index.
- This showed other accounts which had used similar sentences.

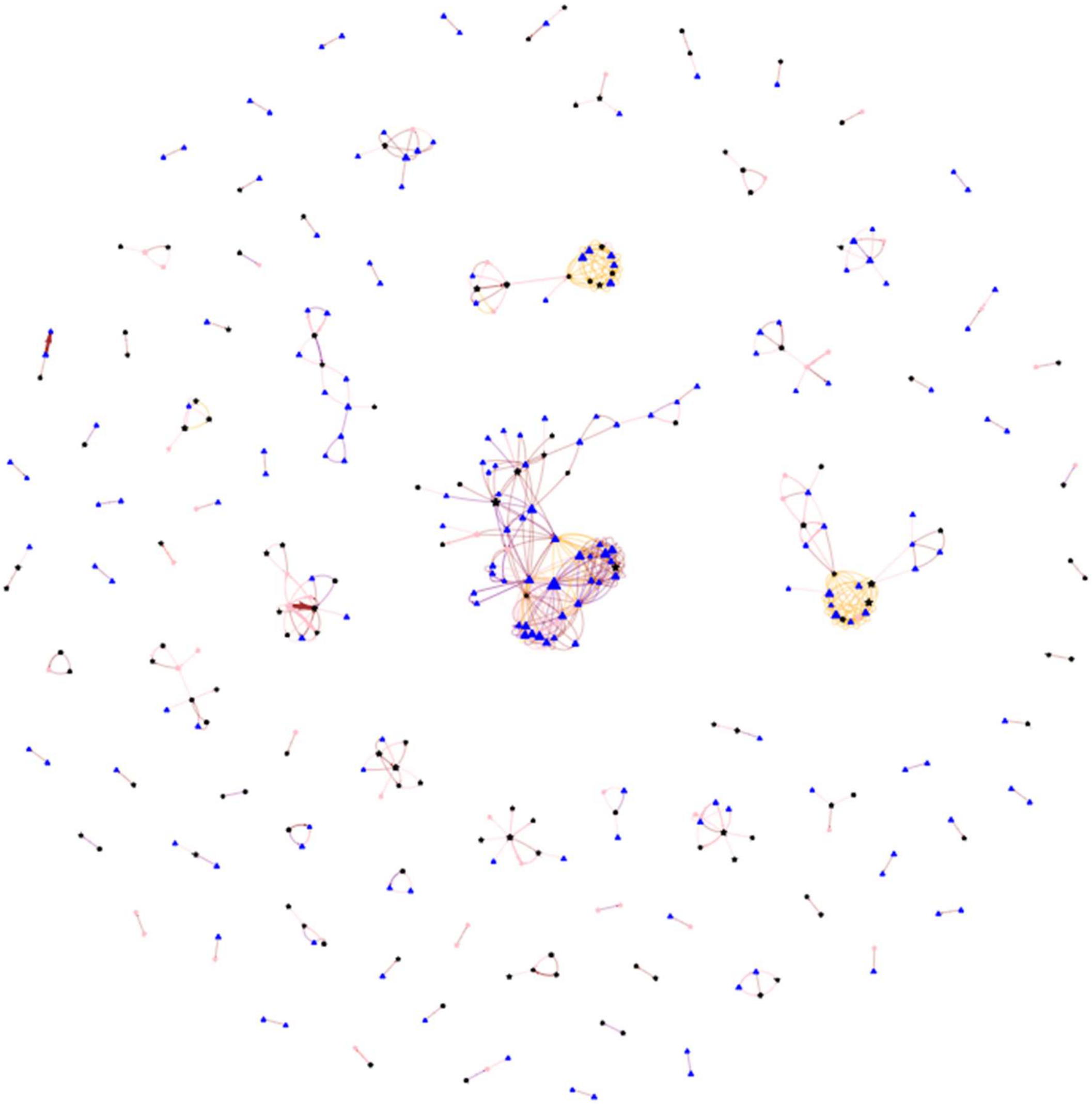

*Figure 8 All results from searches 1-6. There remains noise (from search 4), but some pairs of accounts found in that search are now small clusters of 3 or more accounts. The central cluster (believed to represent a paper mill) is also enlarged by search 6.*

### *Visualisation*

Finally, the data was visualized such that referee-accounts are represented as nodes and 2 referees are connected by an edge if they have produced duplicate comments at any time. Edges connecting referees to authors show where a referee had refereed one or more of that author's papers (at any time – not just those times when a duplicate was found).

False positives were found at this stage where

- Individuals who had refereed for a large number of authors were typically journal editors or board-members. Occasionally board-members will share template comments, or quote parts of referees' comments. These individuals were removed manually from the data at this stage.
- Some duplication of data had occurred when 'practice' documents were created for demonstration and training purposes. These documents were removed from the data.

Summary data was generated to show which nodes have the most connections and which have the fewest. Where possible, these outlier cases were inspected manually and a few obvious false-positives were removed. E.g. cases where a referee account had not been properly de-duplicated and the same referee was using the same template comment in 2 or more separate accounts. After removing these false positives, the searches were re-run. The visualization was then used as a tool to explore the duplicate comments found in the system.

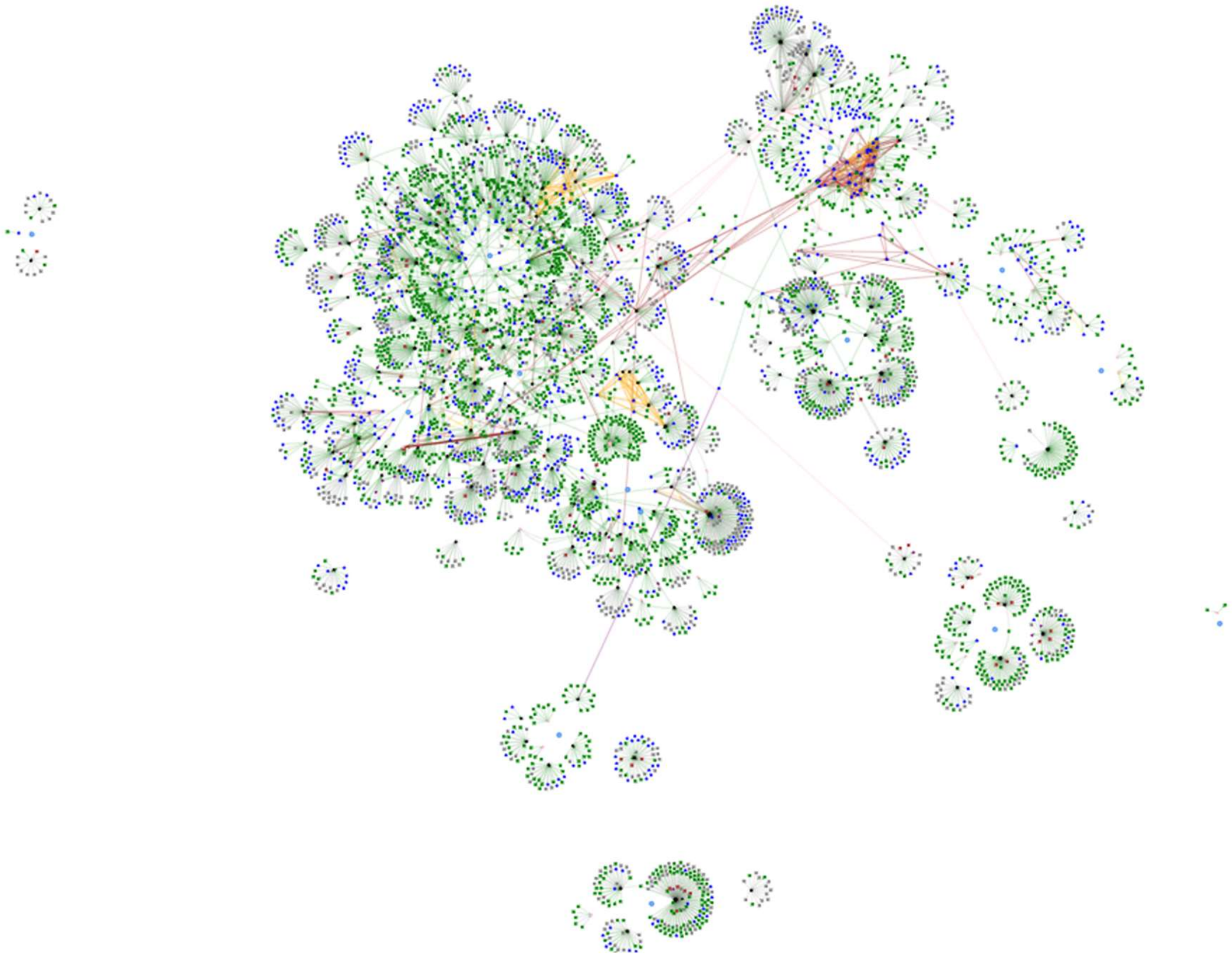

*Graph statistics*

The search methods have detected cases of unusual duplication. However, much of the time, duplication is not cause for concern. Cases which appear to represent peer-review fraud are often cases where we see a lot of links between a group of individuals. We can therefore use graph centrality metrics to rank the individuals in our dataset. Centrality metrics attempt to measure the importance of nodes in a graph. For detailed descriptions of these metrics, see *(Hao Liao, 2017)*.

Here, we use PageRank (Lawrence Page, 1999) to rank the nodes in our graph. This ranking is useful for prioritizing which referee-accounts to investigate for peer-review fraud.

- The graph is represented as a dataframe of referee-account-pairs where each row of the dataframe contains a pair of referee accounts found to have produced duplicate – or partial duplicate reports.
- This dataframe is converted into a NetworkX graph and PageRank is calculated using the built-in `pagerank` function in NetworkX (NetworkX Developers, 2021)

| UID | PageRank |
| --- | --- |
| uid167189 | 0.052397 |
| uid139523 | 0.049650 |
| uid128508 | 0.026730 |
| uid143987 | 0.026192 |
| uid143029 | 0.021690 |
| uid190444 | 0.018388 |
| uid80496 | 0.017180 |
| uid85704 | 0.016008 |

| UID | PageRank |
|---|---|
| uid145841 | 0.015925 |

Table 3 Example of PageRank rankings of referee-accounts

- Once this ranking has been created, pseudonymized UIDs can be un-pseudonymised for the purpose of investigating the referee-account for peer-review fraud.

## Results

Below is summary data showing the rate of duplication in referee-comments. The rate of duplication is very low.

| | |
|---|---|
| No. unique review accounts | 62794 |
| No. unique review accounts removed due to innocent duplication | 29 |
| No. unique review accounts in paper-mill cluster | 47 |
| No. unique review accounts that produced duplicates or partial duplicates | 357 |

Table 4 Percentage of unique reviewers found to have written duplicate comment. These numbers vary depending on the search parameters used.

| | |
|---|---|
| No. articles | 66815 |
| No. articles with reviews from cluster review accounts | 77 |
| No. articles with reviews from any account that produced duplicates | 972 |

Table 5 Percentage of articles which received duplicate comments

There a number of reasons why the data presented above should not be taken as a *general* measure of misconduct rates:

- Data is drawn from a limited (and therefore biased) dataset.
- There is a lot of innocent duplication in referee comments.

- Cutoffs for significance of results can be changed easily and so, all of these numbers can change significantly with a change of search parameters.

However, this may give a rough indication of duplication rates to be expected if similar searches are carried out elsewhere.

| Search | Time to index | Time to search | # accounts found |
|---|---|---|---|
| Search1 | 16s | <1s | 29 |
| Search2 | 13m 52s | <1s | 58 |
| Search3 | 23m 36s | 23m 9s | 98 |
| Search4 | 18m 29s | 9h 19m 2s | 204 |
| Search5 | ~ | <1s | 58 |
| Search6 | 3h 30m 20s | 20m 40s | 297 |

*Table 6 speed comparison for different searches. Search5 has no automated 'indexing' step, but requires manual work to identify common errors in the dataset. Given the number of unique referee accounts found and the scalability of Elasticsearch, searches 4 and 6 appear the most useful.*

This table shows a comparison of the different methods used to detect duplication in peer-review. Each row and column shows the overlap in detections by each method.

| | **Search1** | **Search2** | **Search3** | **Search4** | **Search5** | **Search6** |
|---|---|---|---|---|---|---|
| **Search1** | 0 | 13 | 4 | 5 | 11 | 2 |
| **Search2** | 56 | 0 | 40 | 26 | 52 | 5 |
| **Search3** | 73 | 66 | 0 | 28 | 76 | 17 |
| **Search4** | 180 | 158 | 134 | 0 | 177 | 32 |
| **Search5** | 40 | 38 | 36 | 31 | 0 | 12 |
| **Search6** | 270 | 230 | 216 | 125 | 251 | 0 |

*Table 7 Comparison of results of different searches. Each row shows how many unique ids were found to have produced duplicate, or similar, review comments by each method that were not found by other methods. So search1 found 13 referee accounts which produced similar reviews that search2 did not find and search2 found 56 accounts which search1 did not find. It's remarkable that Search1 (a simple search for exact matches) found some results that other methods (which are intended to find exact matches as well as partial matches) did not detect. This is due to additional pre-processing steps on Search1 that were not used in other searches.*

### *Paper-mill activity*

Making the assumption that the large cluster of referee accounts found to be producing similar comments are controlled by a paper-mill, we find that reports received from these individuals rose sharply in early 2018 and appears to have gradually declined since then.

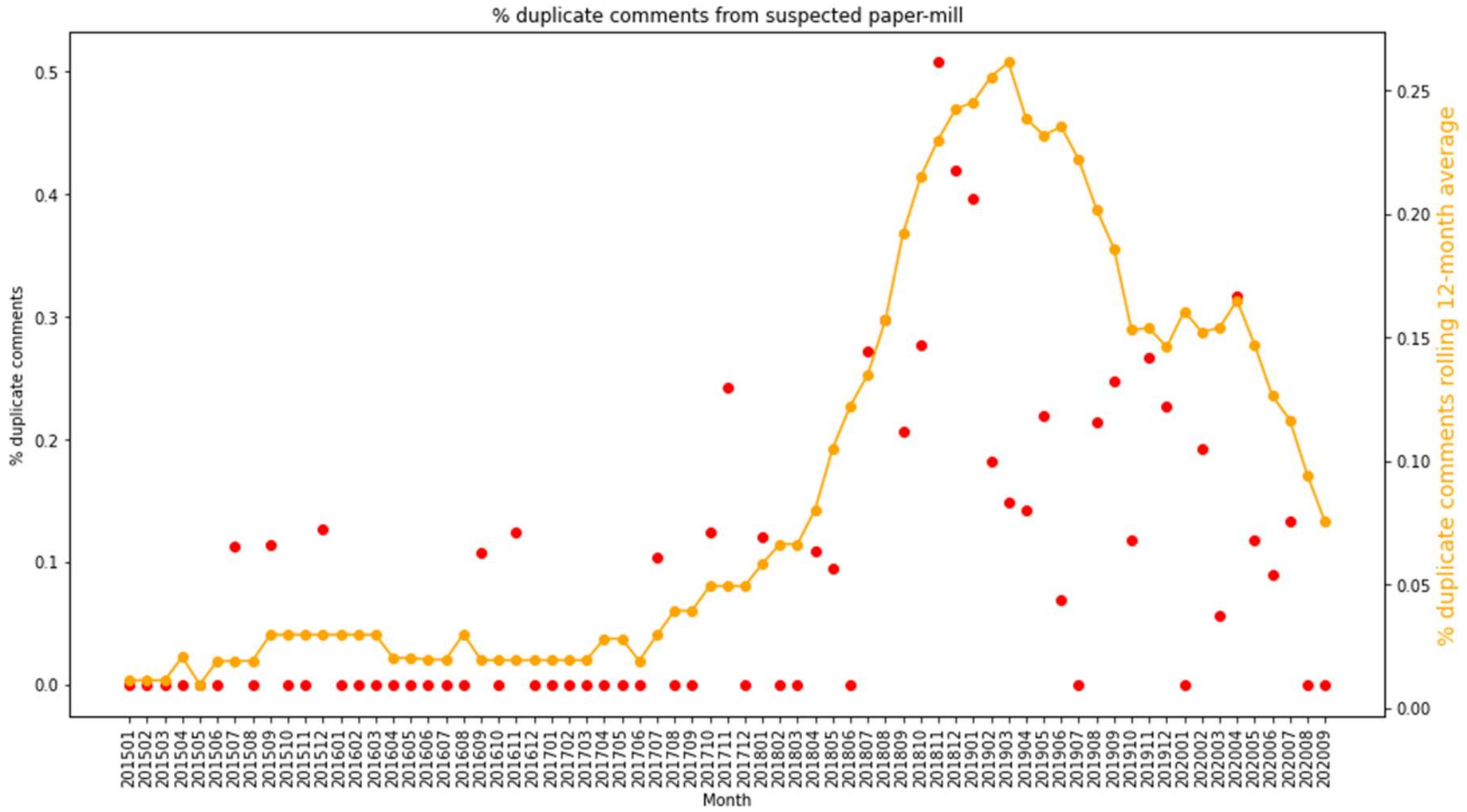

*Figure 9 Percentage of comments from accounts believed to be associated with a paper-mill*

De-pseudonymising the suspect referee-accounts above makes it straightforward to find all of the author accounts which recommended those reviewer-accounts as referees. While the number of suspect referee accounts is low, the number of authors to use them is comparatively high (see Figure 10). So, identifying just 1 fake referee-account can enable the identification of multiple paper-mill submissions.

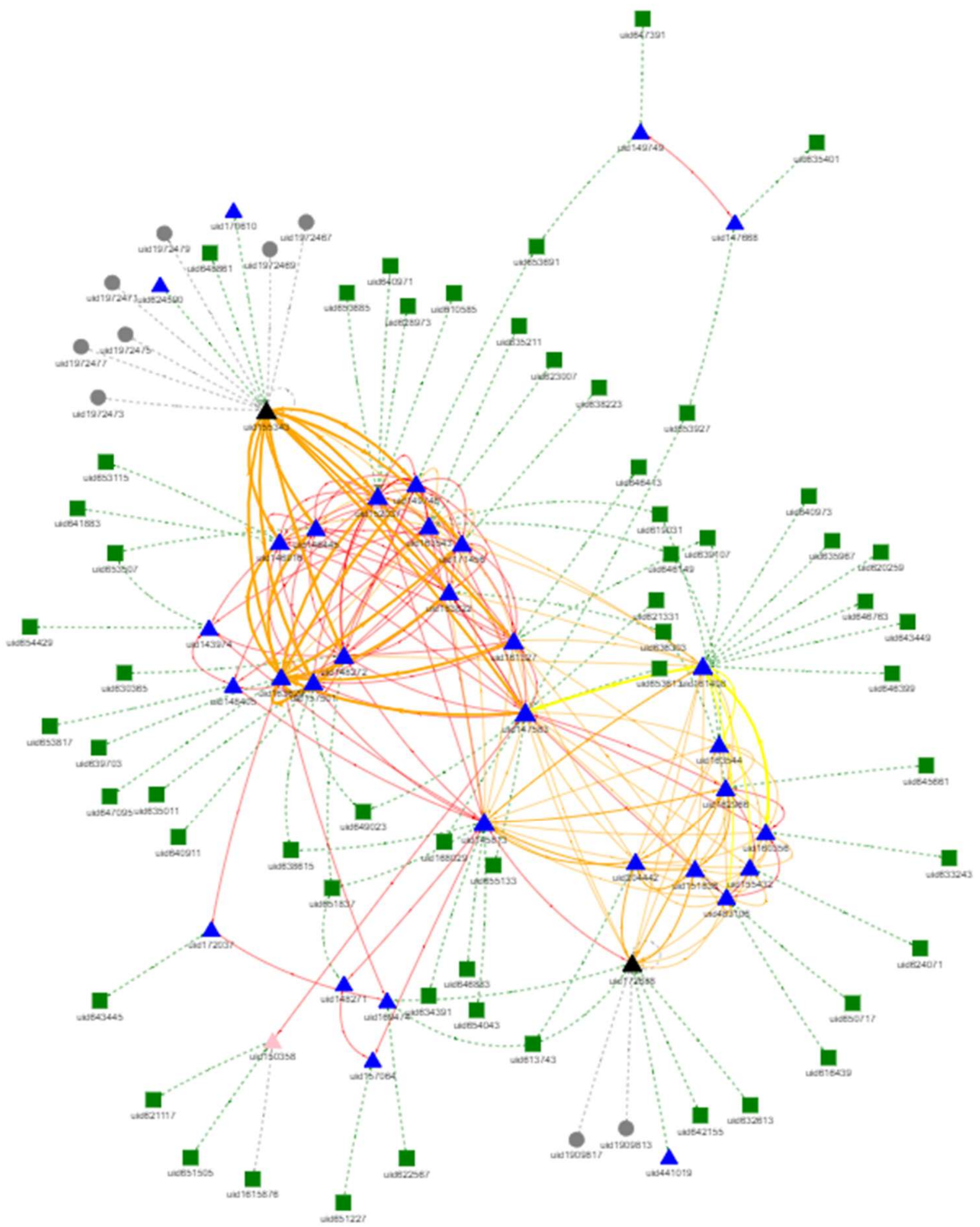

*Figure 10 A cluster of referee accounts (triangles) which have produced similar comments. Authors for those referee accounts are shown in green. In cases where the author recommended one of these referee accounts as a reviewer, it seems likely that the manuscript under consideration was produced by a paper mill.*

## Discussion

A number of methods were used in this study to detect and quantify duplication of text in peer-review comments. Interestingly, none of these methods stand out as being best in terms of speed or quality of results. In a production setting, however, a simple index of hashed comments could be used to find exact duplicates. This could be combined with Elasticsearch indexes (shown in Seaches 4 & 6) to help identify less-obvious cases. These approaches are fast and scalable and would retrieve most of the cases found by the 6 searches presented here.

The methods used in this study are simple and basic. Sadly, this means that peer-review fraudsters can simply ensure that all of their comments are unique in order to avoid being caught by these methods. Indeed, these methods may have failed to detect peer-review fraud where such precautions are already being taken.

Duplication of referee comments is also not proof, in itself, of misconduct. Referees share report templates, and journals issue templates, for innocent reasons. Referees might also copy, or quote, each other's comments at times. However, such innocent duplication is found to be rare and further evidence of misconduct can often be found when the peer-review process is revisited in detail E.g. when the authors have recommended the referee who submitted the duplicate comment.

In recent years, fraudulent research being produced industrially by paper-mills has become a growing problem (Byrne & Christopher, 2020). We speculate that this trend in scientific misconduct is also responsible for an increase in the number of suspicious referee accounts found with the methods presented in this paper (see Figure 9). Detecting peer-review fraud may therefore be one way to detect paper-mill activity.

Some fake referee accounts are created when an author of an article submits a paper and recommends fake account details as a suitable reviewer of their paper. One possible way to avoid peer-review fraud is therefore to simply not allow authors to recommend referees for their work. However:

- a paper-mill might submit a manuscript to a journal under a fake author-account.
- If that author-account is not identified as being fake, there is the risk of that the account will be invited as a referee in the future.

This means that fake referee accounts controlled by paper-mills may be asked to review papers irrespective of whether the author has recommended them.

## Future work

This study was exploratory in nature and, as such, does not quantify rates of peer-review fraud – only rates of duplication which, while unusual, is not necessarily a sign of misconduct. However, there are a number of methods of detecting fraud and unusual activity. Combining these approaches could lead to effective misconduct detection methods. In cases where misconduct has been investigated and diagnosed, it may be possible to build training data for machine-learning approaches to misconduct detection.

In the meantime, treating the problem of peer-review fraud as one of *plagiarism of referee-comments* is a sound approach and anti-plagiarism software, such as Turnitin (Turnitin, 2021) could be valuable for detecting peer-review fraud. However, it is worth considering the prevalence of paraphrasing tools which support plagiarism (Prentice & Kinden, 2018) and which reduce the effectiveness of traditional plagiarism detection software. While there has been much recent progress in paraphrase detection, it remains an unsolved problem in NLP research.

Recent work shows that referee-comments can be easily automatically generated using deep-learning methods. (Bartoli, De Lorenzo, Medvet, & Tarlao, 2016) (Yuan, Liu, & Neubig, 2021). Another recent paper shows one way that such automatically-generated text can be detected (Cabanac & Labbé, 2020).
Future work should examine whether automatically generated peer-review can be detected.

## Conclusion

We have presented a number of simple methods to help identify and quantify peer-review fraud. These methods can be easily applied to any dataset of peer-review comments to help find cases of peer-review fraud.

Elasticsearch-based methods, while slow to index, have the highest success-rate at identifying duplication and are easily applied with limited computational resources. However, brute-force methods are also fast and effective with small datasets. The minHash LSH method finds a middle ground between the two, but requires a lot of random-access memory (RAM) for large datasets.

We find a very small percentage of referee-accounts that have produced duplicate comments. In some cases, it appears obvious that these referee-accounts are controlled by a paper-mill (an organisation which is deliberately seeking to influence the peer-review process). Other cases appear to represent peer-review fraud organised by individuals or smaller groups. In order to confirm the existence of peer-review fraud, however, it will be necessary to investigate each case carefully.

### Acknowledgements

The author is grateful to SAGE colleagues: Helen King, Julie Svalastog and Bailey Baumann for assistance and to Spencer Westby for supplying the data used in this study.

Thanks also to Yury Kashnitsky for recommending the MinHash LSH method used in Search3 and to Retraction Watch, Elisabeth Bik and Guillaume Cabanac for Twitter comments which were helpful in finding relevant literature.